\UseRawInputEncoding
\documentclass[aps,prl,twocolumn,showpacs,superscriptaddress,amsmath,floatfix]{revtex4} 

\usepackage{natbib}
\usepackage[T1]{fontenc}
\usepackage[colorlinks,hyperindex,colorlinks,citecolor=blue,linkcolor=blue,urlcolor=blue,]{hyperref}   
\usepackage{cleveref}
\usepackage{array}
\usepackage{tabularx}
\usepackage{graphicx}  
\graphicspath{{./}}
\usepackage{subfigure}
\usepackage{epstopdf}
\usepackage{dcolumn}        
\usepackage{amssymb} 
\usepackage{amsmath}
\usepackage{booktabs}
\usepackage{multirow}
\usepackage{color}
\usepackage{upgreek}
\usepackage{textcomp}
\usepackage[marginal]{footmisc}

\usepackage{siunitx}

\makeindex

\begin{document}

\title{Atomistic metrics of BaSO$_4$ as an ultra-efficient radiative cooling material: a first-principles prediction}

\author{Zhen Tong}
\thanks{These authors contribute equally to this work.}
\affiliation{School of Mechanical Engineering and the Birck Nanotechnology Center, Purdue University, West Lafayette, Indiana 47907-2088, USA.}

\affiliation{University of Michigan-Shanghai Jiao Tong University Joint Institute, Shanghai Jiao Tong University, Shanghai 200240, China.}

\affiliation{Shenzhen JL Computational Science and Applied Research Institute, Shenzhen 518110,
China.}

\author{Joseph Peoples}
\thanks{These authors contribute equally to this work.}
\affiliation{School of Mechanical Engineering and the Birck Nanotechnology Center, Purdue University, West Lafayette, Indiana 47907-2088, USA.}

\author{Xiangyu Li}
\affiliation{School of Mechanical Engineering and the Birck Nanotechnology Center, Purdue University, West Lafayette, Indiana 47907-2088, USA.}

\author{Xiaolong Yang}
\affiliation{School of Mechanical Engineering and the Birck Nanotechnology Center, Purdue University, West Lafayette, Indiana 47907-2088, USA.}
\affiliation{Institute for Advanced Study, Shenzhen University, Nanhai Avenue 3688, Shenzhen 518060, China.}

\author{Hua Bao}
\email{hua.bao@sjtu.edu.cn}
\affiliation{University of Michigan-Shanghai Jiao Tong University Joint Institute, Shanghai Jiao Tong University, Shanghai 200240, China.}

\author{Xiulin Ruan}
\email{ruan@purdue.edu}
\affiliation{School of Mechanical Engineering and the Birck Nanotechnology Center, Purdue University, West Lafayette, Indiana 47907-2088, USA.}

\date{\today}
	
\begin{abstract}
Radiative cooling has recently revived due to its significant potential as an environmentally friendly cooling technology. However, the design of particle-matrix cooling nanocomposites was generally carried out via tedious trial-and-error approaches, and the atomistic physics for efficient radiative cooling was not well understood. In this work, we identify the atomistic metrics of Barium Sulfate (BaSO$_4$) nanocomposite, which is an ultra-efficient radiative cooling material, using a predictive first-principles approach coupled with Monte Carlo simulations. Our results show that BaSO$_4$-acrylic nanocomposites not only attain high total solar reflectance of 92.5\% (0.28 - 4.0 $\upmu$m), but also simultaneously demonstrate high normal emittance of 96.0\% in the sky window region (8 - 13 $\upmu$m), outperforming the commonly used $\alpha$-quartz ($\alpha$-SiO$_2$). We identify two pertinent characters of ultra-efficient radiative cooling paints: i) a balanced band gap and refractive index, which enables strong scattering while negating absorption in the solar spectrum, and ii) a sufficient number of infrared-active optical resonance phonon modes resulting in abundant Reststrahlen bands and high emissivity in the sky window. The first principles approach and the resulted physical insights in this work pave the way for further search of ultra-efficient radiative cooling materials.

\end{abstract}

\maketitle

Radiative cooling is an important passive cooling technology, and numerous materials have been extensively explored for cooling of buildings \cite{michell_radiation_1979}, solar cells \cite{zhu_radiative_2014}, power plants \cite{zeyghami_performance_2015}, dew water harvesting \cite{liu_hydrophilic_2020}, and desalination \cite{xu_all-day_2020}, etc. Ideal materials for radiative cooling should have high reflectance (low emissivity) in the solar spectrum and high emissivity in the sky window region \cite{family_materials_2017,sun_radiative_2017,zeyghami_review_2018,zhao_radiative_2019}. As such, designing radiative cooling materials or devices with tailored spectral properties has attracted significant research interest. Since 1960s, radiative cooling had been pursued \cite{catalanotti_radiative_1975,harrison_radiative_1978,orel_radiative_1993,pockett_heat_2010}, and a TiO$_2$ paint-aluminum substrate dual-layer showed daytime below-ambient cooling during a winter day \cite{harrison_radiative_1978}. In the last decade, the interest in radiative cooling was renewed \cite{gentle_radiative_2010}. Full daytime radiative cooling was achieved in nanophotonic multilayers \cite{raman_passive_2014}, and subsequently in other non-paint approaches including polymer-metal dual layer \cite{gentle_subambient_2015}, silica nanocomposite-metal dual layer \cite{zhai_scalable-manufactured_2017}, silica-metal dual layer \cite{kou_daytime_2017}, and delignified wood \cite{li_radiative_2019}. Meanwhile, efficient radiative cooling paints have been designed \cite{huang_nanoparticle_2017, bao_double-layer_2017, peoples_strategy_2019}, and full daytime cooling has been demonstrated in porous polymers \cite{mandal_hierarchically_2018} and very recently in commercial-like CaCO$_3$-acrylic and BaSO$_4$-acrylic paints \cite{ruan_wo2020072818_2020, li_full_2020}, providing a viable radiative cooling solution. There is still much room to search for more efficient radiative cooling paints, however it has been generally carried out based on empirical approaches, either by modeling material with known optical constants or experimenting materials lacking the knowledge of optical constants on a trial-and-error basis. The atomistic structural metrics that lead to ultra-efficient radiative cooling paints such as the BaSO$_4$-acrylic paint \cite{ruan_wo2020072818_2020} were not well understood. Predictive approaches and physical insights are needed to facilitate the design of better radiative cooling paints in the future. In fact, first principles calculations of thermal radiative properties have been rare overall.

In this letter, we illustrate the effectiveness of combining first-principles calculations and Monte Carlo simulations to predict and understand the high performance of radiative cooling materials. We quantitatively explain why BaSO$_4$-based nanocomposite, whose optical constants are not available in literature, is a better radiative cooling material than SiO$_2$-based nanocomposite that is the state-of-the-art owing to its wide availability and abundant emission peaks in the sky window region \cite{family_materials_2017,zhao_radiative_2019,gentle_radiative_2010}. The dielectric function of BaSO$_4$ in both solar spectrum and mid-infrared (mid-IR) region is determined from first-principles, where the Bethe-Salpeter equation \cite{rohlfing_electron-hole_1998,albrecht_ab_1998} and anharmonic lattice dynamics are employed for accurately describing the photon-electron interactions \cite{chang_excitons_2000,bao_ab_2010,kresse_optical_2012,yang_temperature-dependent_2014} and photon-lattice absorptions \cite{tong_temperature-dependent_2018,tong_first-principles_2020,yang_observation_2020}, respectively. The dielectric function is then used in Monte Carlo simulations to obtain the transmittance, reflectance, and emittance (or absorptance) of nanoparticle composites \cite{modest_radiative_2013,howell_thermal_2016,wang_monte_1992,redmond_multiple_2004,huang_nanoparticle_2017,peoples_strategy_2019}. Our analysis show that BaSO$_4$ has a balanced band gap and refractive index, which enable strong scattering while negating absorption in the solar spectrum. Moreover, its complex crystal structure and appropriate bond strength give a high number of infrared-active zone-center optical phonon modes in the Reststrahlen bands, which yield high emissivity in the sky window. Our approach and the identified atomistic metrics can serve as effective tools to design efficient radiative cooling paints in the future.    

\emph{First principles theory}.\textthreequartersemdash We have implemented a first-principles approach to obtain the dielectric function. In the solar spectrum, photons are mainly absorbed through electronic transition. By summing all the possible vertical transitions across the band gap, the imaginary part of the dielectric function $\varepsilon(\omega)$ can be determined based on the Fermi's golden rule \cite{zhang_nanomicroscale_2007,gajdos_linear_2006}
\begin{equation}
\label{eq_1}
\begin{split}
\varepsilon_{\alpha\beta}^{''}(\omega)=
&\frac{4\pi^2e^2}{\Omega}\lim_{q\rightarrow0}\frac{1}{q^2} \sum_{c,v,\bf{k}}2w_{\bf{k}}\delta\left(\epsilon_{c\bf{k}}-\epsilon_{v\bf{k}}-\omega \right)\\
&\times\left\langle u_{c\bf{k}+\bf{e}_{\alpha}q}\mid u_{v\bf{k}} \right\rangle\left\langle u_{c\bf{k}+\bf{e}_{\beta}q}\mid u_{v\bf{k}} \right\rangle, 
\end{split} 
\end{equation}
where $\bf{e}_{\alpha}$ is the unit vector for the $\alpha$ direction, $\Omega$ is the unit-cell volume, $w_{\bf{k}}$ is the weight corresponding to the k-point $\bf{k}$, $\epsilon$ is the energy of electron, and $\omega$ is the frequency. The indices $c$ and $v$ denote the conduction and valance band, respectively. The $u_{c\bf{k}}$ is the cell periodic part of the orbitals at the k-point $\bf{k}$. The real part of the dielectric function can be further obtained from $\varepsilon_{\alpha\beta}^{''}(\omega)$ through Kramers-Kr\"{o}nig transformation \cite{zhang_nanomicroscale_2007}
\begin{equation}
\label{eq_2}
\varepsilon_{\alpha\beta}^{'}(\omega)=1+\frac{2}{\pi}P\int_{0}^{\infty} \frac{\varepsilon_{\alpha\beta}^{''}(\omega')\omega'}{\omega'^2-\omega^2}d\omega',
\end{equation}
where $P$ is the principle value of the integral. 

In the mid-IR band, the dielectric function is described by the Lorentz oscillator model \cite{barker_transverse_1964,schubert_coordinate-invariant_2016}
\begin{equation}
\label{eq_dielectric}
\frac{\varepsilon(\omega)}{\varepsilon_{\infty}}=\prod_{j}\frac{\omega_{j,\text{LO}}^2-\omega^2+\text{i}\gamma_{j,\text{LO}}\omega}{\omega_{j,\text{TO}}^2-\omega^2+\text{i}\gamma_{j,\text{TO}}\omega},
\end{equation}
where $\varepsilon_{\infty}$ is the high frequency dielectric constant, $\omega_j$ is the resonance frequency, $j$ goes over all the IR-active modes, and $\gamma_j$ is the damping factor of the $j$-th resonance phonon mode. LO and TO denote the longitudinal and transverse optical phonon modes, respectively. By employing anharmonic lattice dynamics method, $\gamma$ (or scattering rate $\tau^{-1}$) can be calculated using the harmonic and anharmonic interatomic force constants from first-principles calculations \cite{tong_temperature-dependent_2018,tong_first-principles_2020,yang_observation_2020}. 

All first-principles calculations in this work were performed by employing projector-augmented-wave (PAW) \cite{kresse_ultrasoft_1999} method, as implemented in the Vienna ab initio simulation package (VASP) \cite{kresse_ab_1993}, and the Monte Carlo simulations were implemented with open-source code \cite{peoples_strategy_2019}. More details on the computational methodology, structure optimization, force constants calculations, and Monte Carlo computing are provided in Sec. S1 of the Supplemental Material (SM). Importantly, in order to validate our calculations, we first apply our methodology to $\alpha$-$\text{SiO}_2$ which is a benchmark radiative cooling material with sufficient existing experimental data.

\begin{figure}[htp]
	\centerline{\includegraphics[width= 8.6cm]{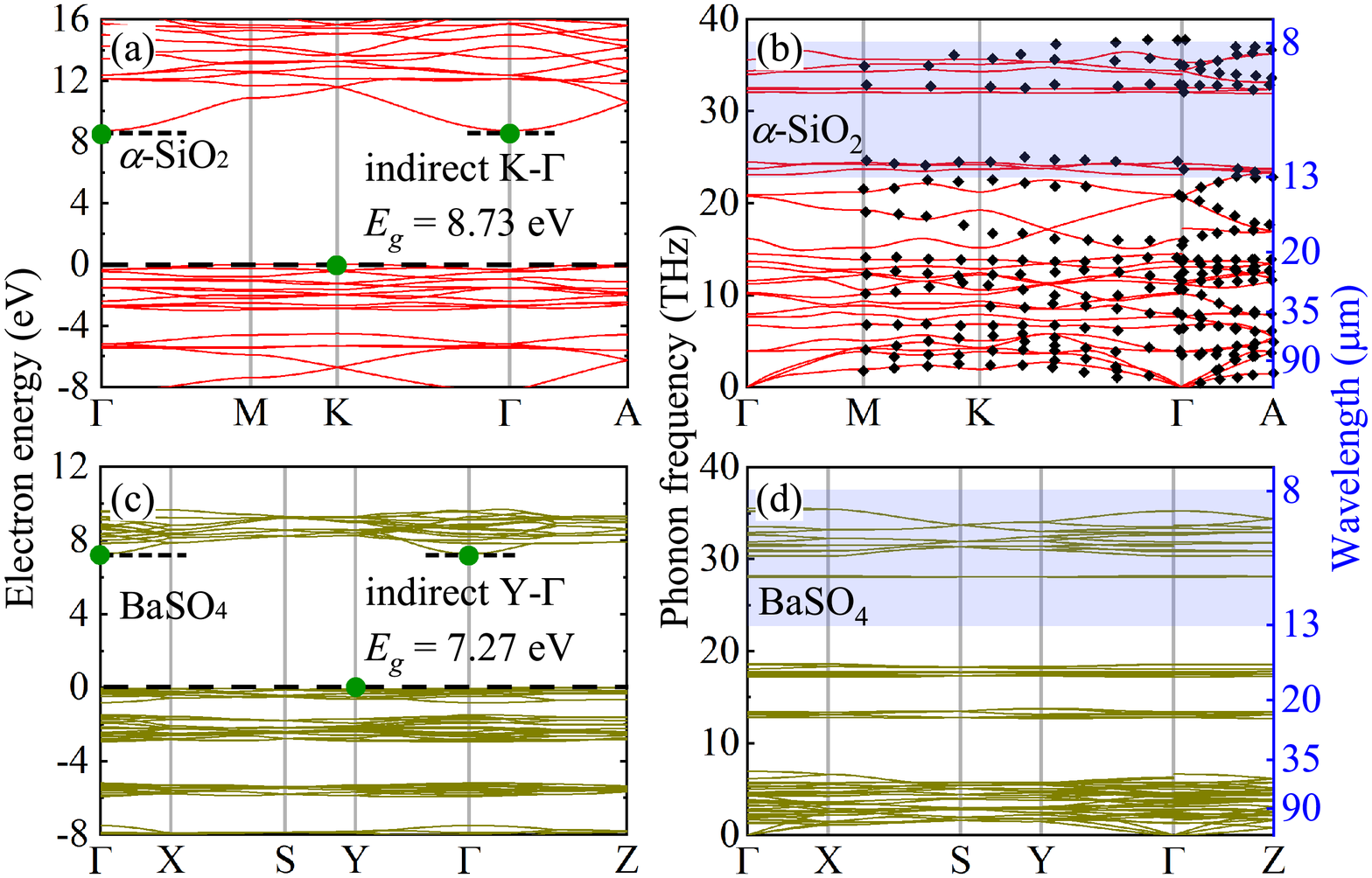}}
	\caption{\label{fig:1}
	Calculated band structure (a) and phonon dispersion (b) for $\alpha$-$\text{SiO}_{2}$ from first-principles, and (c) and (d) for $\text{BaSO}_{4}$, respectively. Experimentally measured \cite{strauch_lattice_1993} phonon frequency (black diamond) of $\alpha$-SiO$_2$ is also presented in (b) for comparison. Indirect band gaps are denoted using green dot at high symmetry point in the first Brillouin zone with K-$\Gamma$ in (a) and Y-$\Gamma$ in (c) for $\alpha$-$\text{SiO}_{2}$ and $\text{BaSO}_{4}$, respectively. The wavelength corresponding to phonon frequency is presented using blue right-$y$ axis, and the sky window ranging in 8-13 $\upmu$m is depicted by shaded region with blue color in (b) and (d), for $\alpha$-$\text{SiO}_{2}$ and $\text{BaSO}_{4}$, respectively.}
\end{figure}

The band structure and phonon dispersion are shown in Figs. \ref{fig:1} (a) and (b) for $\alpha$-$\text{SiO}_{2}$, and (c) and (d) for $\text{BaSO}_{4}$, respectively. Due to the electronic transition from valence band to conduction band when absorbing photon in the solar spectrum, band gap plays a critical role in accurately describing the wavelength dependent dielectric response. When the ground state density functional theory (DFT) is used (such as the traditional generalized gradient, GGA \cite{perdew_generalized_1996}, employed here), the band gap was identified to be 5.67 eV for $\alpha$-$\text{SiO}_{2}$ (previously reported DFT value was 5.80 eV \cite{alkauskas_defect_2008}) and 5.94 eV for $\text{BaSO}_{4}$ (recently published DFT value was 6.03 eV \cite{korabelnikov_structural_2018}). However, we note that ground state DFT generally fails to predict the band gap accurately due to the single particle approximation \cite{alkauskas_defect_2008,kresse_optical_2012}. A more accurate approach of the excited states is the self-consistent quasiparticle GW (sc-QPGW) calculation \cite{kresse_optical_2012} which includes the corrections for the electrostatic interactions between electrons and holes. After conducting sc-QPGW calculations \cite{kresse_optical_2012}, the band gaps were updated to 8.73 eV for $\alpha$-$\text{SiO}_{2}$ (measured value was 8.90 eV \cite{alkauskas_defect_2008}) and 7.27 eV for $\text{BaSO}_{4}$ (experimental value was 7.60 eV \cite{amiryan_recombination_1977}), which were also depicted in their band structure as shown in Figs. \ref{fig:1} (a) and (c), respectively. The prediction-measurement agreement was excellent, laying foundation of the following resulted dielectric function calculations. Here, we observe that the band gaps of both $\alpha$-$\text{SiO}_{2}$ and $\text{BaSO}_{4}$ are higher than the energy of photons located in the solar spectrum with a range of 0.49 $\sim$ 4.13 eV (0.3 - 2.5 $\upmu$m). As such, it is not surprising there is no photon absorption in the solar spectrum for both $\alpha$-$\text{SiO}_{2}$ and $\text{BaSO}_{4}$. 

Furthermore, phonon parameters are calculated to obtain the dielectric response in the mid-IR band. Phonon dispersions along high symmetry path in the first Brillouin zone are shown in Figs. \ref{fig:1} (b) and (d) for $\alpha$-$\text{SiO}_{2}$ and $\text{BaSO}_{4}$, respectively. Previously reported phonon frequencies \cite{strauch_lattice_1993} of $\alpha$-$\text{SiO}_{2}$ are also presented in Fig. \ref{fig:1} (b), and are in good agreement with our predictions. A key insight is that we observed more optical modes in $\text{BaSO}_{4}$ than that of $\alpha$-$\text{SiO}_{2}$ in the sky window region (8 - 13 $\upmu$m) as depicted with blue shaded regions in Figs. \ref{fig:1} (b) and (d), thus providing absorption of a broader wavelength range of IR photons in $\text{BaSO}_{4}$. Moreover, according to space-group analysis \cite{kroumova_bilbao_2003,glazer_vibrate_2009}, the irreducible representations of zone-center ($\bf{q}$=0) optical phonon modes are denoted with 4A$_1$+4A$_2$+8E for $\alpha$-$\text{SiO}_{2}$ and 11A$_g$+7B$_{1g}$+11B$_{2g}$+7B$_{3g}$+7A$_{u}$+10B$_{1u}$+6B$_{2u}$+10B$_{3u}$ for BaSO$_4$, in which the IR-active modes are 4A$_2$+8E for $\alpha$-$\text{SiO}_{2}$ while 10B$_{1u}$+6B$_{2u}$+10B$_{3u}$ for BaSO$_4$, verifying that BaSO$_4$ owns more IR-active phonon modes than that of $\alpha$-$\text{SiO}_{2}$. Further details on identifying the values of phonon frequency and linewidth of the IR-active modes used to resolve mid-IR dielectric function in Eq. \ref{eq_dielectric} are given in Sec. S2 in the SM.

\begin{figure}[htp]
	\centering
	\includegraphics[width= 8.6cm]{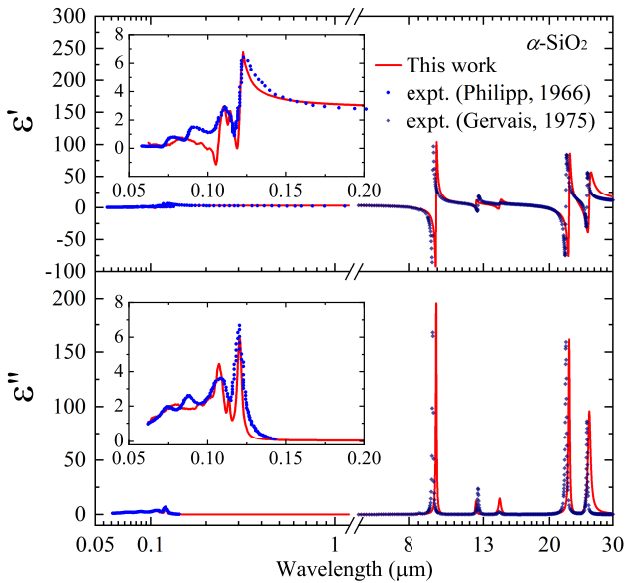}
	\caption{Dielectric function of $\alpha$-$\text{SiO}_{2}$ for wavelength ranging in 0.05-30 $\upmu$m at room temperature. The experimental data in the ultraviolet-visible band are taken from Ref. \cite{philipp_optical_1966} (blue circle dot) while those in the mid-IR region are from Ref. \cite{gervais_temperature_1975} (navy diamond dot).} 
	\label{fig:2}
\end{figure}

Validating our approach is necessary for predictive design of radiative cooling materials without experimental data. As shown in Fig. \ref{fig:2}, for $\alpha$-SiO$_2$, good agreements of the dielectric function between our calculations and reported data in both solar spectrum \cite{philipp_optical_1966} and mid-IR region \cite{gervais_temperature_1975} are achieved, demonstrating the accuracy of our predictions. Therefore, it is reasonable to employ the approach to $\text{BaSO}_4$ or other materials.   

\begin{figure}[htp]
	\includegraphics[width= 8.6cm]{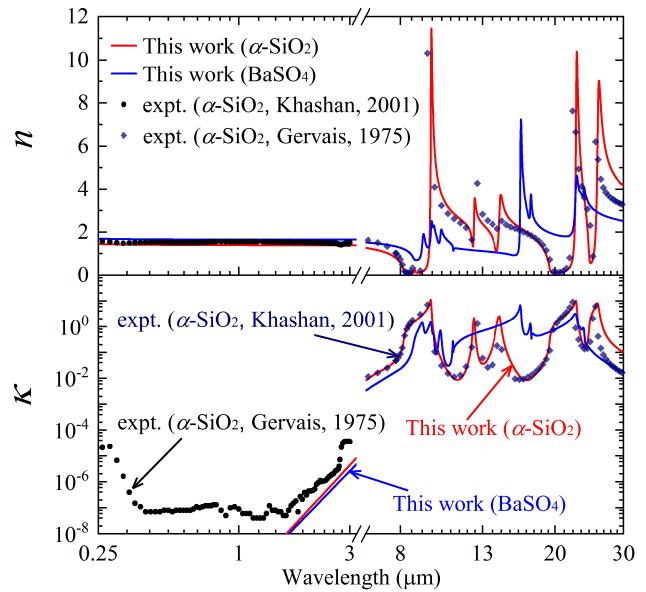}
	\caption{\label{fig:3}
	$n$ and $\kappa$ of $\alpha$-$\text{SiO}_{2}$ and $\text{BaSO}_4$. The experimental data of $\alpha$-$\text{SiO}_{2}$ in both solar spectrum \cite{khashan_dispersion_2001} (black circle dot) and mid-IR region \cite{gervais_temperature_1975} (navy diamond dot) are plotted for comparison.}
\end{figure}

The real and imaginary part of the complex refractive index, $n$ and $\kappa$, can be obtained through $\varepsilon=(n+i\kappa)^2$. The calculated $n$ and $\kappa$ for $\alpha$-$\text{SiO}_{2}$ and $\text{BaSO}_4$ are shown in Fig. \ref{fig:3}. As expected, our predictions on $\alpha$-$\text{SiO}_{2}$ in general agree well with those reported experimental data \cite{khashan_dispersion_2001,gervais_temperature_1975} as shown in Fig. \ref{fig:3}. The $\kappa$ is zero from prediction while negligibly small from experiments probably due to impurities or defects, both indicating negligible absorptance of solar irradiation. Such extremely low extinction coefficient of $\alpha$-$\text{SiO}_{2}$ is due to its large band gap which leads to virtually no absorption in the solar spectrum.

From Fig. \ref{fig:3}, we note that $\kappa$ of $\text{BaSO}_4$ in the solar spectrum approaches zero, which again originates from the large band gap that far exceeds the high-energy boundary (4.13 eV, or 0.3 $\upmu$m) of the solar spectrum. Generally, in semiconductors the band gap is negatively correlated with the refractive index \cite{ravindra_energy_2007}. Therefore, the value of $n$ of $\text{BaSO}_4$ is slightly larger than that of $\alpha$-$\text{SiO}_{2}$ as shown in Fig. \ref{fig:3}, resulting in a larger refractive index contrast between $\text{BaSO}_4$ and typical polymer matrix that has low refractive index. On the other hand, we find that $\text{BaSO}_4$ has larger $\kappa$ than $\alpha$-$\text{SiO}_{2}$ in the mid-IR bands, indicating that $\text{BaSO}_4$ has stronger emission compared to $\alpha$-$\text{SiO}_{2}$ in the sky window. We attribute it to the fact that $\text{BaSO}_4$ owns more IR-active optical phonon modes than that of $\alpha$-$\text{SiO}_{2}$ in the sky window region as indicated in Figs. \ref{fig:1} (b) and (d), and thereby resulting more Reststrahlen bands \cite{adachi_optical_1999} in $\text{BaSO}_4$. 

\begin{figure}[h]
	\includegraphics[width= 8.6cm]{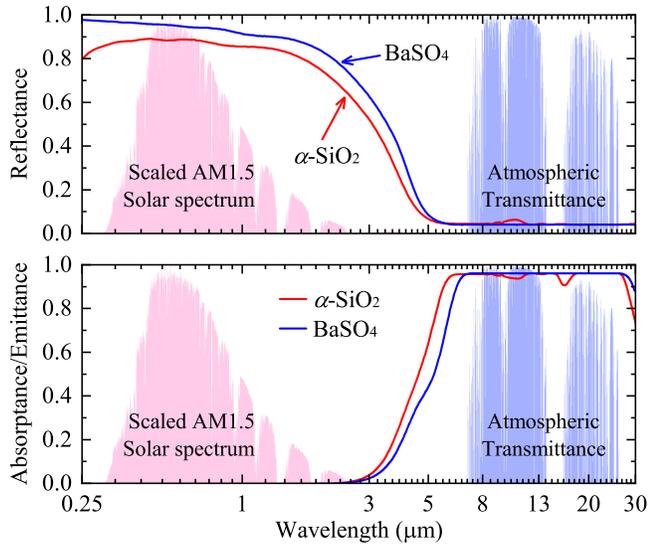}
	\caption{\label{fig:4}
	Reflectance and absorptance/emittance of $\alpha$-$\text{SiO}_{2}$-acrylic (red line) and $\text{BaSO}_4$-acrylic (blue line) nanocomposites calculated from Monte Carlo simulation with particle diameter of 500 nm, volume fraction of 60$\%$, and film thickness of 500 $\upmu$m. The scaled AM1.5 solar spectrum \cite{astm_g159-98_standard_1998} (red area) and the clear sky atmospheric transmittance in the mid-infrared region \cite{lord_new_1992} (blue area) are plotted for reference.}
\end{figure}

Next we investigated the radiative cooling performance in nanocomposite systems where the micro/nano-particulates are embedded in polymer. After employing Mie-theory \cite{redmond_multiple_2004, bohren_absorption_1983, peoples_strategy_2019} to obtain the radiative properties of single particles, Monte Carlo simulations are performed to further obtain the absorptance/emittance, reflectance, and transmittance of nanocomposites consisting of BaSO$_4$ (or SiO$_2$) nanoparticles in an acrylic matrix. It is worth mentioning that we use the above predicted $n$ and $\kappa$ from first-principles as the input parameters to conduct the Monte Carlo simulation without any fitting parameters, indicating that our approach can be adopted for predicting radiative cooling properties of materials. A detailed study of the effect of thickness on the absorptance/emittance was conducted and the results are given in Secs. S3 and S4 in the SM. Here, we present the calculated results of $\alpha$-$\text{SiO}_{2}$-acrylic and $\text{BaSO}_4$-acrylic nanocomposites with particle diameter of 500 nm, volume fraction of 60$\%$, and film thickness of 500 $\upmu$m as shown in Fig. \ref{fig:4}. Importantly, the total solar reflectance was evaluated in the range of 0.28 - 4 $\upmu$m, with the final value of 85.9\% for $\alpha$-$\text{SiO}_{2}$ while 92.5\% for $\text{BaSO}_4$, indicating $\text{BaSO}_4$ outperforms $\alpha$-$\text{SiO}_{2}$ as a radiative cooling material. Moreover, the reflectance of 92.5\% is higher than the typical value of $\sim$ 85\% of commercial white paints \cite{pockett_heat_2010}. Further enhancement of the reflectance can be achieved by using multiple particle sizes \cite{peoples_strategy_2019,li_full_2020}.

In the solar spectrum, due to a higher $n$ resulting from smaller band gap, $\text{BaSO}_4$ shows larger reflectance compared to $\alpha$-$\text{SiO}_{2}$, thereby further corroborates that the BaSO$_4$-based nanocomposites has less transmission of solar irradiation than that of $\alpha$-SiO$_2$. In the sky window, the spectral absorptance or emittance of $\text{BaSO}_4$ is 0.96, which is overall larger than the value 0.94 of $\alpha$-$\text{SiO}_{2}$. This can be attributed to a higher $\kappa$ originating from more IR-active phonon modes in $\text{BaSO}_4$ in the sky window. Both features are desirable for radiative cooling, making $\text{BaSO}_4$ a more efficient radiative cooling material compared to the state-of-the-art $\alpha$-$\text{SiO}_{2}$. The solar reflectance and sky window emissivity together yield a radiative cooling standard figure of merit \cite{li_full_2020} of 0.21, indicating that our $\text{BaSO}_4$-acrylic paint design can be cooled below the ambient temperature under direct sunlight. 

Comparing properties of $\text{BaSO}_4$ and $\alpha$-SiO$_2$ allows us to identify the atomistic metrics for efficient radiative cooling materials. First, radiative cooling materials require high reflection in the solar spectrum, which originates from strong scattering and minimal absorption (i.e. large albedo) of nanoparticles. Therefore, the material should have a balanced band gap and refractive index. The band gap should be larger than 4.13 eV (high-energy boundary of solar spectrum), which avoids absorption of solar irradiation. The refractive index should also be large enough to enhance scattering of light. Since the band gap is usually negatively correlated to refractive index \cite{ravindra_energy_2007}, one should search materials with band gap value that is modestly higher than 4.13 eV. The high thermal emittance in the sky window region requires numerous IR-active phonon modes in the 8 - 13 $\upmu$m. Identifying such metrics is imperative for screening efficient radiative cooling materials. We suspect better radiative cooling performance than BaSO$_4$ is possible in materials with a band gap high than 4.13 eV but lower than that of BaSO$_4$. Moreover, when coupled with high-throughput search methods, our approach can efficiently identify the best radiative cooling materials among the large number of candidates in the future.  

To conclude, we demonstrated a multiscale simulation strategy combining first-principles calculation and Monte Carlo simulation to predict the radiative cooling properties of $\alpha$-$\text{SiO}_{2}$- and $\text{BaSO}_4$-based nanocomposites, and it opens the door for predictive design of radiative cooling materials from atomistic simulations. We validate our first-principles calculations on $\alpha$-$\text{SiO}_{2}$ which is a state-of-the-art radiative cooling material. We find that $\text{BaSO}_4$-acrylic nanocomposite has both higher solar reflectance and sky window emittance than that of $\alpha$-$\text{SiO}_{2}$, demonstrating it to be a more efficient radiative cooling material compared to $\alpha$-$\text{SiO}_{2}$. More importantly, we identified two significant atomistic metrics for ultra-efficient radiative cooling materials: a balanced band gap and refractive index, which induce larger scattering and negligible absorption in the solar spectrum; and a sufficient number of infrared-active optical resonance phonon modes, which lead to abundant Reststrahlen bands and thereby offering high emissivity in the sky window region. This work provides insights on how material structures and atomistic metrics can determine intrinsic properties and lead to the design and/or screening of ultra-efficient radiative cooling materials. Other radiative applications could benefit from the first principles approach and insights as well, including thermal barrier coatings, concentrated solar receiver materials, thermophotovoltaic coatings, and variable emissivitiy coatings for space vehicles.

\begin{center}
\textbf{ACKNOWLEDGMENTS}
\end{center}

Simulations were preformed at the Rosen Center for Advanced Computing (RCAC) of Purdue University. J.P., X.L., and X.R. thank the Cooling Technologies Research Center (CTRC) at Purdue University for partial support of this work. H.B. acknowledges the support by the National Natural Science Foundation of China (Grant No. 51676121). Z.T. acknowledges the financial support of Chinese Scholarship Council (CSC, No. 201806230169), China Postdoctoral Science Foundation (Grant No. 2020M680127), and Fundamental and Applied Fundamental Research Foundation of Guangdong Province (Grant No. 2020A1515110838).

\bibliographystyle{apsrev4_updates}
\bibliography{Reference}

\begin{thebibliography}{59}%
\makeatletter
\providecommand \@ifxundefined [1]{%
 \@ifx{#1\undefined}
}%
\providecommand \@ifnum [1]{%
 \ifnum #1\expandafter \@firstoftwo
 \else \expandafter \@secondoftwo
 \fi
}%
\providecommand \@ifx [1]{%
 \ifx #1\expandafter \@firstoftwo
 \else \expandafter \@secondoftwo
 \fi
}%
\providecommand \natexlab [1]{#1}%
\providecommand \enquote  [1]{``#1''}%
\providecommand \bibnamefont  [1]{#1}%
\providecommand \bibfnamefont [1]{#1}%
\providecommand \citenamefont [1]{#1}%
\providecommand \href@noop [0]{\@secondoftwo}%
\providecommand \href [0]{\begingroup \@sanitize@url \@href}%
\providecommand \@href[1]{\@@startlink{#1}\@@href}%
\providecommand \@@href[1]{\endgroup#1\@@endlink}%
\providecommand \@sanitize@url [0]{\catcode `\\12\catcode `\$12\catcode
  `\&12\catcode `\#12\catcode `\^12\catcode `\_12\catcode `\%12\relax}%
\providecommand \@@startlink[1]{}%
\providecommand \@@endlink[0]{}%
\providecommand \url  [0]{\begingroup\@sanitize@url \@url }%
\providecommand \@url [1]{\endgroup\@href {#1}{\urlprefix }}%
\providecommand \urlprefix  [0]{URL }%
\providecommand \Eprint [0]{\href }%
\providecommand \doibase [0]{https://doi.org/}%
\providecommand \selectlanguage [0]{\@gobble}%
\providecommand \bibinfo  [0]{\@secondoftwo}%
\providecommand \bibfield  [0]{\@secondoftwo}%
\providecommand \translation [1]{[#1]}%
\providecommand \BibitemOpen [0]{}%
\providecommand \bibitemStop [0]{}%
\providecommand \bibitemNoStop [0]{.\EOS\space}%
\providecommand \EOS [0]{\spacefactor3000\relax}%
\providecommand \BibitemShut  [1]{\csname bibitem#1\endcsname}%
\let\auto@bib@innerbib\@empty
\bibitem [{\citenamefont {Michell}\ and\ \citenamefont
  {Biggs}(1979)}]{michell_radiation_1979}%
  \BibitemOpen
  \bibfield  {author} {\bibinfo {author} {\bibfnamefont {D.}~\bibnamefont
  {Michell}}\ and\ \bibinfo {author} {\bibfnamefont {K.}~\bibnamefont
  {Biggs}},\ }\bibfield  {title} {\emph {\bibinfo {title} {Radiation cooling of
  buildings at night}},\ }\href {https://doi.org/10.1016/0306-2619(79)90017-5}
  {\bibfield  {journal} {\bibinfo  {journal} {Applied Energy}\ }\textbf
  {\bibinfo {volume} {5}},\ \bibinfo {pages} {263} (\bibinfo {year}
  {1979})}\BibitemShut {NoStop}%
\bibitem [{\citenamefont {Zhu}\ \emph {et~al.}(2014)\citenamefont {Zhu},
  \citenamefont {Raman}, \citenamefont {Wang}, \citenamefont {Anoma},\ and\
  \citenamefont {Fan}}]{zhu_radiative_2014}%
  \BibitemOpen
  \bibfield  {author} {\bibinfo {author} {\bibfnamefont {L.}~\bibnamefont
  {Zhu}}, \bibinfo {author} {\bibfnamefont {A.}~\bibnamefont {Raman}}, \bibinfo
  {author} {\bibfnamefont {K.~X.}\ \bibnamefont {Wang}}, \bibinfo {author}
  {\bibfnamefont {M.~A.}\ \bibnamefont {Anoma}},\ and\ \bibinfo {author}
  {\bibfnamefont {S.}~\bibnamefont {Fan}},\ }\bibfield  {title} {\emph
  {\bibinfo {title} {Radiative cooling of solar cells}},\ }\href
  {https://doi.org/10.1364/OPTICA.1.000032} {\bibfield  {journal} {\bibinfo
  {journal} {Optica}\ }\textbf {\bibinfo {volume} {1}},\ \bibinfo {pages} {32}
  (\bibinfo {year} {2014})}\BibitemShut {NoStop}%
\bibitem [{\citenamefont {Zeyghami}\ and\ \citenamefont
  {Khalili}(2015)}]{zeyghami_performance_2015}%
  \BibitemOpen
  \bibfield  {author} {\bibinfo {author} {\bibfnamefont {M.}~\bibnamefont
  {Zeyghami}}\ and\ \bibinfo {author} {\bibfnamefont {F.}~\bibnamefont
  {Khalili}},\ }\bibfield  {title} {\emph {\bibinfo {title} {Performance
  improvement of dry cooled advanced concentrating solar power plants using
  daytime radiative cooling}},\ }\href
  {https://doi.org/10.1016/j.enconman.2015.09.016} {\bibfield  {journal}
  {\bibinfo  {journal} {Energy Conversion and Management}\ }\textbf {\bibinfo
  {volume} {106}},\ \bibinfo {pages} {10} (\bibinfo {year} {2015})}\BibitemShut
  {NoStop}%
\bibitem [{\citenamefont {Liu}\ \emph {et~al.}(2020)\citenamefont {Liu},
  \citenamefont {Fan},\ and\ \citenamefont {Bao}}]{liu_hydrophilic_2020}%
  \BibitemOpen
  \bibfield  {author} {\bibinfo {author} {\bibfnamefont {C.}~\bibnamefont
  {Liu}}, \bibinfo {author} {\bibfnamefont {J.}~\bibnamefont {Fan}},\ and\
  \bibinfo {author} {\bibfnamefont {H.}~\bibnamefont {Bao}},\ }\bibfield
  {title} {\emph {\bibinfo {title} {Hydrophilic radiative cooler for direct
  water condensation in humid weather}},\ }\href
  {https://doi.org/10.1016/j.solmat.2020.110700} {\bibfield  {journal}
  {\bibinfo  {journal} {Solar Energy Materials \& Solar Cells}\ }\textbf
  {\bibinfo {volume} {216}},\ \bibinfo {pages} {110700} (\bibinfo {year}
  {2020})}\BibitemShut {NoStop}%
\bibitem [{\citenamefont {Xu}\ \emph {et~al.}(2020)\citenamefont {Xu},
  \citenamefont {Zhang}, \citenamefont {Fu}, \citenamefont {Song},
  \citenamefont {Shang}, \citenamefont {Tao},\ and\ \citenamefont
  {Deng}}]{xu_all-day_2020}%
  \BibitemOpen
  \bibfield  {author} {\bibinfo {author} {\bibfnamefont {J.}~\bibnamefont
  {Xu}}, \bibinfo {author} {\bibfnamefont {J.}~\bibnamefont {Zhang}}, \bibinfo
  {author} {\bibfnamefont {B.}~\bibnamefont {Fu}}, \bibinfo {author}
  {\bibfnamefont {C.}~\bibnamefont {Song}}, \bibinfo {author} {\bibfnamefont
  {W.}~\bibnamefont {Shang}}, \bibinfo {author} {\bibfnamefont
  {P.}~\bibnamefont {Tao}},\ and\ \bibinfo {author} {\bibfnamefont
  {T.}~\bibnamefont {Deng}},\ }\bibfield  {title} {\emph {\bibinfo {title}
  {All-{Day} {Freshwater} {Harvesting} through {Combined} {Solar}-{Driven}
  {Interfacial} {Desalination} and {Passive} {Radiative} {Cooling}}},\ }\href
  {https://pubs.acs.org/doi/abs/10.1021/acsami.0c14773} {\bibfield  {journal}
  {\bibinfo  {journal} {ACS Applied Materials \& Interfaces}\ }\textbf
  {\bibinfo {volume} {12}},\ \bibinfo {pages} {47612} (\bibinfo {year}
  {2020})}\BibitemShut {NoStop}%
\bibitem [{\citenamefont {Family}\ and\ \citenamefont
  {Meng\"{u}\c{c}}(2017)}]{family_materials_2017}%
  \BibitemOpen
  \bibfield  {author} {\bibinfo {author} {\bibfnamefont {R.}~\bibnamefont
  {Family}}\ and\ \bibinfo {author} {\bibfnamefont {M.~P.}\ \bibnamefont
  {Meng\"{u}\c{c}}},\ }\bibfield  {title} {\emph {\bibinfo {title} {Materials
  for {Radiative} {Cooling}: {A} {Review}}},\ }\href
  {https://doi.org/10.1016/j.proenv.2017.03.158} {\bibfield  {journal}
  {\bibinfo  {journal} {Procedia Environmental Sciences}\ }\textbf {\bibinfo
  {volume} {38}},\ \bibinfo {pages} {752} (\bibinfo {year} {2017})}\BibitemShut
  {NoStop}%
\bibitem [{\citenamefont {Sun}\ \emph {et~al.}(2017)\citenamefont {Sun},
  \citenamefont {Sun}, \citenamefont {Zhou}, \citenamefont {Alam},\ and\
  \citenamefont {Bermel}}]{sun_radiative_2017}%
  \BibitemOpen
  \bibfield  {author} {\bibinfo {author} {\bibfnamefont {X.}~\bibnamefont
  {Sun}}, \bibinfo {author} {\bibfnamefont {Y.}~\bibnamefont {Sun}}, \bibinfo
  {author} {\bibfnamefont {Z.}~\bibnamefont {Zhou}}, \bibinfo {author}
  {\bibfnamefont {M.~A.}\ \bibnamefont {Alam}},\ and\ \bibinfo {author}
  {\bibfnamefont {P.}~\bibnamefont {Bermel}},\ }\bibfield  {title} {\emph
  {\bibinfo {title} {Radiative sky cooling: fundamental physics, materials,
  structures, and applications}},\ }\href
  {https://doi.org/10.1515/nanoph-2017-0020} {\bibfield  {journal} {\bibinfo
  {journal} {Nanophotonics}\ }\textbf {\bibinfo {volume} {6}},\ \bibinfo
  {pages} {997} (\bibinfo {year} {2017})}\BibitemShut {NoStop}%
\bibitem [{\citenamefont {Zeyghami}\ \emph {et~al.}(2018)\citenamefont
  {Zeyghami}, \citenamefont {Goswami},\ and\ \citenamefont
  {Stefanakos}}]{zeyghami_review_2018}%
  \BibitemOpen
  \bibfield  {author} {\bibinfo {author} {\bibfnamefont {M.}~\bibnamefont
  {Zeyghami}}, \bibinfo {author} {\bibfnamefont {D.~Y.}\ \bibnamefont
  {Goswami}},\ and\ \bibinfo {author} {\bibfnamefont {E.}~\bibnamefont
  {Stefanakos}},\ }\bibfield  {title} {\emph {\bibinfo {title} {A review of
  clear sky radiative cooling developments and applications in renewable power
  systems and passive building cooling}},\ }\href
  {https://doi.org/10.1016/j.solmat.2018.01.015} {\bibfield  {journal}
  {\bibinfo  {journal} {Solar Energy Materials and Solar Cells}\ }\textbf
  {\bibinfo {volume} {178}},\ \bibinfo {pages} {115} (\bibinfo {year}
  {2018})}\BibitemShut {NoStop}%
\bibitem [{\citenamefont {Zhao}\ \emph {et~al.}(2019)\citenamefont {Zhao},
  \citenamefont {Aili}, \citenamefont {Zhai}, \citenamefont {Xu}, \citenamefont
  {Tan}, \citenamefont {Yin},\ and\ \citenamefont
  {Yang}}]{zhao_radiative_2019}%
  \BibitemOpen
  \bibfield  {author} {\bibinfo {author} {\bibfnamefont {D.}~\bibnamefont
  {Zhao}}, \bibinfo {author} {\bibfnamefont {A.}~\bibnamefont {Aili}}, \bibinfo
  {author} {\bibfnamefont {Y.}~\bibnamefont {Zhai}}, \bibinfo {author}
  {\bibfnamefont {S.}~\bibnamefont {Xu}}, \bibinfo {author} {\bibfnamefont
  {G.}~\bibnamefont {Tan}}, \bibinfo {author} {\bibfnamefont {X.}~\bibnamefont
  {Yin}},\ and\ \bibinfo {author} {\bibfnamefont {R.}~\bibnamefont {Yang}},\
  }\bibfield  {title} {\emph {\bibinfo {title} {Radiative sky cooling:
  {Fundamental} principles, materials, and applications}},\ }\href
  {https://doi.org/10.1063/1.5087281} {\bibfield  {journal} {\bibinfo
  {journal} {Applied Physics Reviews}\ }\textbf {\bibinfo {volume} {6}},\
  \bibinfo {pages} {021306} (\bibinfo {year} {2019})}\BibitemShut {NoStop}%
\bibitem [{\citenamefont {Catalanotti}\ \emph {et~al.}(1975)\citenamefont
  {Catalanotti}, \citenamefont {Cuomo}, \citenamefont {Piro}, \citenamefont
  {Ruggi}, \citenamefont {Silvestrini},\ and\ \citenamefont
  {Troise}}]{catalanotti_radiative_1975}%
  \BibitemOpen
  \bibfield  {author} {\bibinfo {author} {\bibfnamefont {S.}~\bibnamefont
  {Catalanotti}}, \bibinfo {author} {\bibfnamefont {V.}~\bibnamefont {Cuomo}},
  \bibinfo {author} {\bibfnamefont {G.}~\bibnamefont {Piro}}, \bibinfo {author}
  {\bibfnamefont {D.}~\bibnamefont {Ruggi}}, \bibinfo {author} {\bibfnamefont
  {V.}~\bibnamefont {Silvestrini}},\ and\ \bibinfo {author} {\bibfnamefont
  {G.}~\bibnamefont {Troise}},\ }\bibfield  {title} {\emph {\bibinfo {title}
  {The radiative cooling of selective surfaces}},\ }\href
  {https://doi.org/10.1016/0038-092X(75)90062-6} {\bibfield  {journal}
  {\bibinfo  {journal} {Solar Energy}\ }\textbf {\bibinfo {volume} {17}},\
  \bibinfo {pages} {83} (\bibinfo {year} {1975})}\BibitemShut {NoStop}%
\bibitem [{\citenamefont {Harrison}\ and\ \citenamefont
  {Walton}(1978)}]{harrison_radiative_1978}%
  \BibitemOpen
  \bibfield  {author} {\bibinfo {author} {\bibfnamefont {A.}~\bibnamefont
  {Harrison}}\ and\ \bibinfo {author} {\bibfnamefont {M.}~\bibnamefont
  {Walton}},\ }\bibfield  {title} {\emph {\bibinfo {title} {Radiative cooling
  of {Ti}{O}$_{\textrm{2}}$ white paint}},\ }\href
  {https://doi.org/10.1016/0038-092X(78)90195-0} {\bibfield  {journal}
  {\bibinfo  {journal} {Solar Energy}\ }\textbf {\bibinfo {volume} {20}},\
  \bibinfo {pages} {185} (\bibinfo {year} {1978})}\BibitemShut {NoStop}%
\bibitem [{\citenamefont {Orel}\ \emph {et~al.}(1993)\citenamefont {Orel},
  \citenamefont {Gunde},\ and\ \citenamefont {Krainer}}]{orel_radiative_1993}%
  \BibitemOpen
  \bibfield  {author} {\bibinfo {author} {\bibfnamefont {B.}~\bibnamefont
  {Orel}}, \bibinfo {author} {\bibfnamefont {M.~K.}\ \bibnamefont {Gunde}},\
  and\ \bibinfo {author} {\bibfnamefont {A.}~\bibnamefont {Krainer}},\
  }\bibfield  {title} {\emph {\bibinfo {title} {Radiative cooling efficiency of
  white pigmented paints}},\ }\href
  {https://doi.org/10.1016/0038-092X(93)90108-Z} {\bibfield  {journal}
  {\bibinfo  {journal} {Solar Energy}\ }\textbf {\bibinfo {volume} {50}},\
  \bibinfo {pages} {477} (\bibinfo {year} {1993})}\BibitemShut {NoStop}%
\bibitem [{\citenamefont {Pockett}(2010)}]{pockett_heat_2010}%
  \BibitemOpen
  \bibfield  {author} {\bibinfo {author} {\bibfnamefont {J.}~\bibnamefont
  {Pockett}},\ }in\ \href
  {https://search.informit.com.au/documentSummary;dn=995653800330276;res=IELENG}
  {\emph {\bibinfo {booktitle} {Chemeca 2010: {Engineering} at the {Edge}}}}\
  (\bibinfo {address} {Hilton Adelaide, South Australia},\ \bibinfo {year}
  {2010})\ p.\ \bibinfo {pages} {2999},\ \bibinfo {note} {publisher: Engineers
  Australia}\BibitemShut {NoStop}%
\bibitem [{\citenamefont {Gentle}\ and\ \citenamefont
  {Smith}(2010)}]{gentle_radiative_2010}%
  \BibitemOpen
  \bibfield  {author} {\bibinfo {author} {\bibfnamefont {A.~R.}\ \bibnamefont
  {Gentle}}\ and\ \bibinfo {author} {\bibfnamefont {G.~B.}\ \bibnamefont
  {Smith}},\ }\bibfield  {title} {\emph {\bibinfo {title} {Radiative {Heat}
  {Pumping} from the {Earth} {Using} {Surface} {Phonon} {Resonant}
  {Nanoparticles}}},\ }\href {https://doi.org/10.1021/nl903271d} {\bibfield
  {journal} {\bibinfo  {journal} {Nano Letters}\ }\textbf {\bibinfo {volume}
  {10}},\ \bibinfo {pages} {373} (\bibinfo {year} {2010})}\BibitemShut
  {NoStop}%
\bibitem [{\citenamefont {Raman}\ \emph {et~al.}(2014)\citenamefont {Raman},
  \citenamefont {Anoma}, \citenamefont {Zhu}, \citenamefont {Rephaeli},\ and\
  \citenamefont {Fan}}]{raman_passive_2014}%
  \BibitemOpen
  \bibfield  {author} {\bibinfo {author} {\bibfnamefont {A.~P.}\ \bibnamefont
  {Raman}}, \bibinfo {author} {\bibfnamefont {M.~A.}\ \bibnamefont {Anoma}},
  \bibinfo {author} {\bibfnamefont {L.}~\bibnamefont {Zhu}}, \bibinfo {author}
  {\bibfnamefont {E.}~\bibnamefont {Rephaeli}},\ and\ \bibinfo {author}
  {\bibfnamefont {S.}~\bibnamefont {Fan}},\ }\bibfield  {title} {\emph
  {\bibinfo {title} {Passive radiative cooling below ambient air temperature
  under direct sunlight}},\ }\href {https://doi.org/10.1038/nature13883}
  {\bibfield  {journal} {\bibinfo  {journal} {Nature}\ }\textbf {\bibinfo
  {volume} {515}},\ \bibinfo {pages} {540} (\bibinfo {year}
  {2014})}\BibitemShut {NoStop}%
\bibitem [{\citenamefont {Gentle}\ and\ \citenamefont
  {Smith}(2015)}]{gentle_subambient_2015}%
  \BibitemOpen
  \bibfield  {author} {\bibinfo {author} {\bibfnamefont {A.~R.}\ \bibnamefont
  {Gentle}}\ and\ \bibinfo {author} {\bibfnamefont {G.~B.}\ \bibnamefont
  {Smith}},\ }\bibfield  {title} {\emph {\bibinfo {title} {A {Subambient}
  {Open} {Roof} {Surface} under the {Mid}-{Summer} {Sun}}},\ }\href
  {https://doi.org/https://doi.org/10.1002/advs.201500119} {\bibfield
  {journal} {\bibinfo  {journal} {Advanced Science}\ }\textbf {\bibinfo
  {volume} {2}},\ \bibinfo {pages} {1500119} (\bibinfo {year}
  {2015})}\BibitemShut {NoStop}%
\bibitem [{\citenamefont {Zhai}\ \emph {et~al.}(2017)\citenamefont {Zhai},
  \citenamefont {Ma}, \citenamefont {David}, \citenamefont {Zhao},
  \citenamefont {Lou}, \citenamefont {Tan}, \citenamefont {Yang},\ and\
  \citenamefont {Yin}}]{zhai_scalable-manufactured_2017}%
  \BibitemOpen
  \bibfield  {author} {\bibinfo {author} {\bibfnamefont {Y.}~\bibnamefont
  {Zhai}}, \bibinfo {author} {\bibfnamefont {Y.}~\bibnamefont {Ma}}, \bibinfo
  {author} {\bibfnamefont {S.~N.}\ \bibnamefont {David}}, \bibinfo {author}
  {\bibfnamefont {D.}~\bibnamefont {Zhao}}, \bibinfo {author} {\bibfnamefont
  {R.}~\bibnamefont {Lou}}, \bibinfo {author} {\bibfnamefont {G.}~\bibnamefont
  {Tan}}, \bibinfo {author} {\bibfnamefont {R.}~\bibnamefont {Yang}},\ and\
  \bibinfo {author} {\bibfnamefont {X.}~\bibnamefont {Yin}},\ }\bibfield
  {title} {\emph {\bibinfo {title} {Scalable-manufactured randomized
  glass-polymer hybrid metamaterial for daytime radiative cooling}},\ }\href
  {https://doi.org/10.1126/science.aai7899} {\bibfield  {journal} {\bibinfo
  {journal} {Science}\ }\textbf {\bibinfo {volume} {355}},\ \bibinfo {pages}
  {1062} (\bibinfo {year} {2017})}\BibitemShut {NoStop}%
\bibitem [{\citenamefont {Kou}\ \emph {et~al.}(2017)\citenamefont {Kou},
  \citenamefont {Jurado}, \citenamefont {Chen}, \citenamefont {Fan},\ and\
  \citenamefont {Minnich}}]{kou_daytime_2017}%
  \BibitemOpen
  \bibfield  {author} {\bibinfo {author} {\bibfnamefont {J.-l.}\ \bibnamefont
  {Kou}}, \bibinfo {author} {\bibfnamefont {Z.}~\bibnamefont {Jurado}},
  \bibinfo {author} {\bibfnamefont {Z.}~\bibnamefont {Chen}}, \bibinfo {author}
  {\bibfnamefont {S.}~\bibnamefont {Fan}},\ and\ \bibinfo {author}
  {\bibfnamefont {A.~J.}\ \bibnamefont {Minnich}},\ }\bibfield  {title} {\emph
  {\bibinfo {title} {Daytime {Radiative} {Cooling} {Using} {Near}-{Black}
  {Infrared} {Emitters}}},\ }\href
  {https://doi.org/10.1021/acsphotonics.6b00991} {\bibfield  {journal}
  {\bibinfo  {journal} {ACS Photonics}\ }\textbf {\bibinfo {volume} {4}},\
  \bibinfo {pages} {626} (\bibinfo {year} {2017})}\BibitemShut {NoStop}%
\bibitem [{\citenamefont {Li}\ \emph {et~al.}(2019)\citenamefont {Li},
  \citenamefont {Zhai}, \citenamefont {He}, \citenamefont {Gan}, \citenamefont
  {Wei}, \citenamefont {Heidarinejad}, \citenamefont {Dalgo}, \citenamefont
  {Mi}, \citenamefont {Zhao}, \citenamefont {Song}, \citenamefont {Dai},
  \citenamefont {Chen}, \citenamefont {Aili}, \citenamefont {Vellore},
  \citenamefont {Martini}, \citenamefont {Yang}, \citenamefont {Srebric},
  \citenamefont {Yin},\ and\ \citenamefont {Hu}}]{li_radiative_2019}%
  \BibitemOpen
  \bibfield  {author} {\bibinfo {author} {\bibfnamefont {T.}~\bibnamefont
  {Li}}, \bibinfo {author} {\bibfnamefont {Y.}~\bibnamefont {Zhai}}, \bibinfo
  {author} {\bibfnamefont {S.}~\bibnamefont {He}}, \bibinfo {author}
  {\bibfnamefont {W.}~\bibnamefont {Gan}}, \bibinfo {author} {\bibfnamefont
  {Z.}~\bibnamefont {Wei}}, \bibinfo {author} {\bibfnamefont {M.}~\bibnamefont
  {Heidarinejad}}, \bibinfo {author} {\bibfnamefont {D.}~\bibnamefont {Dalgo}},
  \bibinfo {author} {\bibfnamefont {R.}~\bibnamefont {Mi}}, \bibinfo {author}
  {\bibfnamefont {X.}~\bibnamefont {Zhao}}, \bibinfo {author} {\bibfnamefont
  {J.}~\bibnamefont {Song}}, \bibinfo {author} {\bibfnamefont {J.}~\bibnamefont
  {Dai}}, \bibinfo {author} {\bibfnamefont {C.}~\bibnamefont {Chen}}, \bibinfo
  {author} {\bibfnamefont {A.}~\bibnamefont {Aili}}, \bibinfo {author}
  {\bibfnamefont {A.}~\bibnamefont {Vellore}}, \bibinfo {author} {\bibfnamefont
  {A.}~\bibnamefont {Martini}}, \bibinfo {author} {\bibfnamefont
  {R.}~\bibnamefont {Yang}}, \bibinfo {author} {\bibfnamefont {J.}~\bibnamefont
  {Srebric}}, \bibinfo {author} {\bibfnamefont {X.}~\bibnamefont {Yin}},\ and\
  \bibinfo {author} {\bibfnamefont {L.}~\bibnamefont {Hu}},\ }\bibfield
  {title} {\emph {\bibinfo {title} {A radiative cooling structural material}},\
  }\href {https://doi.org/10.1126/science.aau9101} {\bibfield  {journal}
  {\bibinfo  {journal} {Science}\ }\textbf {\bibinfo {volume} {364}},\ \bibinfo
  {pages} {760} (\bibinfo {year} {2019})}\BibitemShut {NoStop}%
\bibitem [{\citenamefont {Huang}\ and\ \citenamefont
  {Ruan}(2017)}]{huang_nanoparticle_2017}%
  \BibitemOpen
  \bibfield  {author} {\bibinfo {author} {\bibfnamefont {Z.}~\bibnamefont
  {Huang}}\ and\ \bibinfo {author} {\bibfnamefont {X.}~\bibnamefont {Ruan}},\
  }\bibfield  {title} {\emph {\bibinfo {title} {Nanoparticle embedded
  double-layer coating for daytime radiative cooling}},\ }\href
  {https://www.sciencedirect.com/science/article/pii/S0017931016309255}
  {\bibfield  {journal} {\bibinfo  {journal} {Int. Journal of Heat and Mass
  Transfer}\ }\textbf {\bibinfo {volume} {104}},\ \bibinfo {pages} {890}
  (\bibinfo {year} {2017})}\BibitemShut {NoStop}%
\bibitem [{\citenamefont {Bao}\ \emph {et~al.}(2017)\citenamefont {Bao},
  \citenamefont {Yan}, \citenamefont {Wang}, \citenamefont {Fang},
  \citenamefont {Zhao},\ and\ \citenamefont {Ruan}}]{bao_double-layer_2017}%
  \BibitemOpen
  \bibfield  {author} {\bibinfo {author} {\bibfnamefont {H.}~\bibnamefont
  {Bao}}, \bibinfo {author} {\bibfnamefont {C.}~\bibnamefont {Yan}}, \bibinfo
  {author} {\bibfnamefont {B.}~\bibnamefont {Wang}}, \bibinfo {author}
  {\bibfnamefont {X.}~\bibnamefont {Fang}}, \bibinfo {author} {\bibfnamefont
  {C.~Y.}\ \bibnamefont {Zhao}},\ and\ \bibinfo {author} {\bibfnamefont
  {X.}~\bibnamefont {Ruan}},\ }\bibfield  {title} {\emph {\bibinfo {title}
  {Double-layer nanoparticle-based coatings for efficient terrestrial radiative
  cooling}},\ }\href {https://doi.org/10.1016/j.solmat.2017.04.020} {\bibfield
  {journal} {\bibinfo  {journal} {Solar Energy Materials and Solar Cells}\
  }\textbf {\bibinfo {volume} {168}},\ \bibinfo {pages} {78} (\bibinfo {year}
  {2017})}\BibitemShut {NoStop}%
\bibitem [{\citenamefont {Peoples}\ \emph {et~al.}(2019)\citenamefont
  {Peoples}, \citenamefont {Li}, \citenamefont {Lv}, \citenamefont {Qiu},
  \citenamefont {Huang},\ and\ \citenamefont {Ruan}}]{peoples_strategy_2019}%
  \BibitemOpen
  \bibfield  {author} {\bibinfo {author} {\bibfnamefont {J.}~\bibnamefont
  {Peoples}}, \bibinfo {author} {\bibfnamefont {X.}~\bibnamefont {Li}},
  \bibinfo {author} {\bibfnamefont {Y.}~\bibnamefont {Lv}}, \bibinfo {author}
  {\bibfnamefont {J.}~\bibnamefont {Qiu}}, \bibinfo {author} {\bibfnamefont
  {Z.}~\bibnamefont {Huang}},\ and\ \bibinfo {author} {\bibfnamefont
  {X.}~\bibnamefont {Ruan}},\ }\bibfield  {title} {\emph {\bibinfo {title} {A
  strategy of hierarchical particle sizes in nanoparticle composite for
  enhancing solar reflection}},\ }\href
  {https://doi.org/10.1016/j.ijheatmasstransfer.2018.11.059} {\bibfield
  {journal} {\bibinfo  {journal} {Int. Journal of Heat and Mass Transfer}\
  }\textbf {\bibinfo {volume} {131}},\ \bibinfo {pages} {487} (\bibinfo {year}
  {2019})}\BibitemShut {NoStop}%
\bibitem [{\citenamefont {Mandal}\ \emph {et~al.}(2018)\citenamefont {Mandal},
  \citenamefont {Fu}, \citenamefont {Overvig}, \citenamefont {Jia},
  \citenamefont {Sun}, \citenamefont {Shi}, \citenamefont {Zhou}, \citenamefont
  {Xiao}, \citenamefont {Yu},\ and\ \citenamefont
  {Yang}}]{mandal_hierarchically_2018}%
  \BibitemOpen
  \bibfield  {author} {\bibinfo {author} {\bibfnamefont {J.}~\bibnamefont
  {Mandal}}, \bibinfo {author} {\bibfnamefont {Y.}~\bibnamefont {Fu}}, \bibinfo
  {author} {\bibfnamefont {A.~C.}\ \bibnamefont {Overvig}}, \bibinfo {author}
  {\bibfnamefont {M.}~\bibnamefont {Jia}}, \bibinfo {author} {\bibfnamefont
  {K.}~\bibnamefont {Sun}}, \bibinfo {author} {\bibfnamefont {N.~N.}\
  \bibnamefont {Shi}}, \bibinfo {author} {\bibfnamefont {H.}~\bibnamefont
  {Zhou}}, \bibinfo {author} {\bibfnamefont {X.}~\bibnamefont {Xiao}}, \bibinfo
  {author} {\bibfnamefont {N.}~\bibnamefont {Yu}},\ and\ \bibinfo {author}
  {\bibfnamefont {Y.}~\bibnamefont {Yang}},\ }\bibfield  {title} {\emph
  {\bibinfo {title} {Hierarchically porous polymer coatings for highly
  efficient passive daytime radiative cooling}},\ }\href
  {https://doi.org/10.1126/science.aat9513} {\bibfield  {journal} {\bibinfo
  {journal} {Science}\ }\textbf {\bibinfo {volume} {362}},\ \bibinfo {pages}
  {315} (\bibinfo {year} {2018})}\BibitemShut {NoStop}%
\bibitem [{\citenamefont {Ruan}\ \emph {et~al.}(2020)\citenamefont {Ruan},
  \citenamefont {Li}, \citenamefont {Huang},\ and\ \citenamefont
  {Peoples}}]{ruan_wo2020072818_2020}%
  \BibitemOpen
  \bibfield  {author} {\bibinfo {author} {\bibfnamefont {X.}~\bibnamefont
  {Ruan}}, \bibinfo {author} {\bibfnamefont {X.}~\bibnamefont {Li}}, \bibinfo
  {author} {\bibfnamefont {Z.}~\bibnamefont {Huang}},\ and\ \bibinfo {author}
  {\bibfnamefont {J.~A.}\ \bibnamefont {Peoples}},\ }\bibfield  {title} {\emph
  {\bibinfo {title} {{WO2020072818} - {Metal}-{Free} {Solar}-{Reflective}
  {Infrared}-{Emissive} {Paints} and {Methods} of {Producing} the {Same}}},\
  }\href {https://patents.google.com/patent/WO2020072818A1/en} {\bibfield
  {journal} {\bibinfo  {journal} {Patent application no. PCT/US2019/054566}\ }
  (\bibinfo {year} {2020})}\BibitemShut {NoStop}%
\bibitem [{\citenamefont {Li}\ \emph {et~al.}(2020)\citenamefont {Li},
  \citenamefont {Peoples}, \citenamefont {Huang}, \citenamefont {Zhao},
  \citenamefont {Qiu},\ and\ \citenamefont {Ruan}}]{li_full_2020}%
  \BibitemOpen
  \bibfield  {author} {\bibinfo {author} {\bibfnamefont {X.}~\bibnamefont
  {Li}}, \bibinfo {author} {\bibfnamefont {J.}~\bibnamefont {Peoples}},
  \bibinfo {author} {\bibfnamefont {Z.}~\bibnamefont {Huang}}, \bibinfo
  {author} {\bibfnamefont {Z.}~\bibnamefont {Zhao}}, \bibinfo {author}
  {\bibfnamefont {J.}~\bibnamefont {Qiu}},\ and\ \bibinfo {author}
  {\bibfnamefont {X.}~\bibnamefont {Ruan}},\ }\bibfield  {title} {\emph
  {\bibinfo {title} {Full {Daytime} {Sub}-ambient {Radiative} {Cooling} in
  {Commercial}-like {Paints} with {High} {Figure} of {Merit}}},\ }\href
  {https://doi.org/10.1016/j.xcrp.2020.100221} {\bibfield  {journal} {\bibinfo
  {journal} {Cell Reports Physical Science}\ }\textbf {\bibinfo {volume} {1}},\
  \bibinfo {pages} {100221} (\bibinfo {year} {2020})}\BibitemShut {NoStop}%
\bibitem [{\citenamefont {Rohlfing}\ and\ \citenamefont
  {Louie}(1998)}]{rohlfing_electron-hole_1998}%
  \BibitemOpen
  \bibfield  {author} {\bibinfo {author} {\bibfnamefont {M.}~\bibnamefont
  {Rohlfing}}\ and\ \bibinfo {author} {\bibfnamefont {S.~G.}\ \bibnamefont
  {Louie}},\ }\bibfield  {title} {\emph {\bibinfo {title} {Electron-{Hole}
  {Excitations} in {Semiconductors} and {Insulators}}},\ }\href
  {https://doi.org/10.1103/PhysRevLett.81.2312} {\bibfield  {journal} {\bibinfo
   {journal} {Physical Review Letters}\ }\textbf {\bibinfo {volume} {81}},\
  \bibinfo {pages} {2312} (\bibinfo {year} {1998})}\BibitemShut {NoStop}%
\bibitem [{\citenamefont {Albrecht}\ \emph {et~al.}(1998)\citenamefont
  {Albrecht}, \citenamefont {Reining}, \citenamefont {Del~Sole},\ and\
  \citenamefont {Onida}}]{albrecht_ab_1998}%
  \BibitemOpen
  \bibfield  {author} {\bibinfo {author} {\bibfnamefont {S.}~\bibnamefont
  {Albrecht}}, \bibinfo {author} {\bibfnamefont {L.}~\bibnamefont {Reining}},
  \bibinfo {author} {\bibfnamefont {R.}~\bibnamefont {Del~Sole}},\ and\
  \bibinfo {author} {\bibfnamefont {G.}~\bibnamefont {Onida}},\ }\bibfield
  {title} {\emph {\bibinfo {title} {\textit{{Ab} {Initio}} {Calculation} of
  {Excitonic} {Effects} in the {Optical} {Spectra} of {Semiconductors}}},\
  }\href {https://doi.org/10.1103/PhysRevLett.80.4510} {\bibfield  {journal}
  {\bibinfo  {journal} {Physical Review Letters}\ }\textbf {\bibinfo {volume}
  {80}},\ \bibinfo {pages} {4510} (\bibinfo {year} {1998})}\BibitemShut
  {NoStop}%
\bibitem [{\citenamefont {Chang}\ \emph {et~al.}(2000)\citenamefont {Chang},
  \citenamefont {Rohlfing},\ and\ \citenamefont {Louie}}]{chang_excitons_2000}%
  \BibitemOpen
  \bibfield  {author} {\bibinfo {author} {\bibfnamefont {E.~K.}\ \bibnamefont
  {Chang}}, \bibinfo {author} {\bibfnamefont {M.}~\bibnamefont {Rohlfing}},\
  and\ \bibinfo {author} {\bibfnamefont {S.~G.}\ \bibnamefont {Louie}},\
  }\bibfield  {title} {\emph {\bibinfo {title} {Excitons and {Optical}
  {Properties} of {$\alpha$}-{Quartz}}},\ }\href
  {https://journals.aps.org/prl/abstract/10.1103/PhysRevLett.85.2613}
  {\bibfield  {journal} {\bibinfo  {journal} {Physical Review Letters}\
  }\textbf {\bibinfo {volume} {85}},\ \bibinfo {pages} {2613} (\bibinfo {year}
  {2000})}\BibitemShut {NoStop}%
\bibitem [{\citenamefont {Bao}\ and\ \citenamefont {Ruan}(2010)}]{bao_ab_2010}%
  \BibitemOpen
  \bibfield  {author} {\bibinfo {author} {\bibfnamefont {H.}~\bibnamefont
  {Bao}}\ and\ \bibinfo {author} {\bibfnamefont {X.}~\bibnamefont {Ruan}},\
  }\bibfield  {title} {\emph {\bibinfo {title} {Ab initio calculations of
  thermal radiative properties: {The} semiconductor {GaAs}}},\ }\href
  {https://doi.org/10.1016/j.ijheatmasstransfer.2009.12.033} {\bibfield
  {journal} {\bibinfo  {journal} {Int. Journal of Heat and Mass Transfer}\
  }\textbf {\bibinfo {volume} {53}},\ \bibinfo {pages} {1308} (\bibinfo {year}
  {2010})}\BibitemShut {NoStop}%
\bibitem [{\citenamefont {Kresse}\ \emph {et~al.}(2012)\citenamefont {Kresse},
  \citenamefont {Marsman}, \citenamefont {Hintzsche},\ and\ \citenamefont
  {Flage-Larsen}}]{kresse_optical_2012}%
  \BibitemOpen
  \bibfield  {author} {\bibinfo {author} {\bibfnamefont {G.}~\bibnamefont
  {Kresse}}, \bibinfo {author} {\bibfnamefont {M.}~\bibnamefont {Marsman}},
  \bibinfo {author} {\bibfnamefont {L.~E.}\ \bibnamefont {Hintzsche}},\ and\
  \bibinfo {author} {\bibfnamefont {E.}~\bibnamefont {Flage-Larsen}},\
  }\bibfield  {title} {\emph {\bibinfo {title} {Optical and electronic
  properties of {Si}$_{\textrm{3}}${N}$_{\textrm{4}}$ and
  {$\alpha$}-{SiO}$_{\textrm{2}}$}},\ }\href
  {https://doi.org/10.1103/PhysRevB.85.045205} {\bibfield  {journal} {\bibinfo
  {journal} {Physical Review B}\ }\textbf {\bibinfo {volume} {85}},\ \bibinfo
  {pages} {045205} (\bibinfo {year} {2012})}\BibitemShut {NoStop}%
\bibitem [{\citenamefont {Yang}\ \emph {et~al.}(2014)\citenamefont {Yang},
  \citenamefont {Liu},\ and\ \citenamefont
  {Tan}}]{yang_temperature-dependent_2014}%
  \BibitemOpen
  \bibfield  {author} {\bibinfo {author} {\bibfnamefont {J.~Y.}\ \bibnamefont
  {Yang}}, \bibinfo {author} {\bibfnamefont {L.~H.}\ \bibnamefont {Liu}},\ and\
  \bibinfo {author} {\bibfnamefont {J.~Y.}\ \bibnamefont {Tan}},\ }\bibfield
  {title} {\emph {\bibinfo {title} {Temperature-dependent dielectric function
  of germanium in the uv-vis spectral range: {A} first-principles study}},\
  }\href {https://doi.org/10.1016/j.jqsrt.2014.02.026} {\bibfield  {journal}
  {\bibinfo  {journal} {J Quant Spectrosc Radiat Transf}\ }\textbf {\bibinfo
  {volume} {141}},\ \bibinfo {pages} {24} (\bibinfo {year} {2014})}\BibitemShut
  {NoStop}%
\bibitem [{\citenamefont {Tong}\ \emph {et~al.}(2018)\citenamefont {Tong},
  \citenamefont {Liu}, \citenamefont {Li},\ and\ \citenamefont
  {Bao}}]{tong_temperature-dependent_2018}%
  \BibitemOpen
  \bibfield  {author} {\bibinfo {author} {\bibfnamefont {Z.}~\bibnamefont
  {Tong}}, \bibinfo {author} {\bibfnamefont {L.}~\bibnamefont {Liu}}, \bibinfo
  {author} {\bibfnamefont {L.}~\bibnamefont {Li}},\ and\ \bibinfo {author}
  {\bibfnamefont {H.}~\bibnamefont {Bao}},\ }\bibfield  {title} {\emph
  {\bibinfo {title} {Temperature-dependent infrared optical properties of
  {3C}-, {4H}- and {6H}-{SiC}}},\ }\href
  {https://doi.org/10.1016/j.physb.2018.02.023} {\bibfield  {journal} {\bibinfo
   {journal} {Physica B: Condensed Matter}\ }\textbf {\bibinfo {volume}
  {537}},\ \bibinfo {pages} {194} (\bibinfo {year} {2018})}\BibitemShut
  {NoStop}%
\bibitem [{\citenamefont {Tong}\ \emph {et~al.}(2020)\citenamefont {Tong},
  \citenamefont {Yang}, \citenamefont {Feng}, \citenamefont {Bao},\ and\
  \citenamefont {Ruan}}]{tong_first-principles_2020}%
  \BibitemOpen
  \bibfield  {author} {\bibinfo {author} {\bibfnamefont {Z.}~\bibnamefont
  {Tong}}, \bibinfo {author} {\bibfnamefont {X.}~\bibnamefont {Yang}}, \bibinfo
  {author} {\bibfnamefont {T.}~\bibnamefont {Feng}}, \bibinfo {author}
  {\bibfnamefont {H.}~\bibnamefont {Bao}},\ and\ \bibinfo {author}
  {\bibfnamefont {X.}~\bibnamefont {Ruan}},\ }\bibfield  {title} {\emph
  {\bibinfo {title} {First-principles predictions of temperature-dependent
  infrared dielectric function of polar materials by including four-phonon
  scattering and phonon frequency shift}},\ }\href
  {https://doi.org/10.1103/PhysRevB.101.125416} {\bibfield  {journal} {\bibinfo
   {journal} {Physical Review B}\ }\textbf {\bibinfo {volume} {101}},\ \bibinfo
  {pages} {125416} (\bibinfo {year} {2020})}\BibitemShut {NoStop}%
\bibitem [{\citenamefont {Yang}\ \emph {et~al.}(2020)\citenamefont {Yang},
  \citenamefont {Feng}, \citenamefont {Kang}, \citenamefont {Hu}, \citenamefont
  {Li},\ and\ \citenamefont {Ruan}}]{yang_observation_2020}%
  \BibitemOpen
  \bibfield  {author} {\bibinfo {author} {\bibfnamefont {X.}~\bibnamefont
  {Yang}}, \bibinfo {author} {\bibfnamefont {T.}~\bibnamefont {Feng}}, \bibinfo
  {author} {\bibfnamefont {J.~S.}\ \bibnamefont {Kang}}, \bibinfo {author}
  {\bibfnamefont {Y.}~\bibnamefont {Hu}}, \bibinfo {author} {\bibfnamefont
  {J.}~\bibnamefont {Li}},\ and\ \bibinfo {author} {\bibfnamefont
  {X.}~\bibnamefont {Ruan}},\ }\bibfield  {title} {\emph {\bibinfo {title}
  {Observation of strong higher-order lattice anharmonicity in {Raman} and
  infrared spectra}},\ }\href {https://doi.org/10.1103/PhysRevB.101.161202}
  {\bibfield  {journal} {\bibinfo  {journal} {Physical Review B}\ }\textbf
  {\bibinfo {volume} {101}},\ \bibinfo {pages} {161202} (\bibinfo {year}
  {2020})}\BibitemShut {NoStop}%
\bibitem [{\citenamefont {Modest}(2013)}]{modest_radiative_2013}%
  \BibitemOpen
  \bibfield  {author} {\bibinfo {author} {\bibfnamefont {M.~F.}\ \bibnamefont
  {Modest}},\ }\href@noop {} {\emph {\bibinfo {title} {Radiative heat
  transfer}}},\ \bibinfo {edition} {3rd}\ ed.\ (\bibinfo  {publisher} {Academic
  Press},\ \bibinfo {address} {New York},\ \bibinfo {year} {2013})\BibitemShut
  {NoStop}%
\bibitem [{\citenamefont {Howell}\ and\ \citenamefont
  {Robert}(2016)}]{howell_thermal_2016}%
  \BibitemOpen
  \bibfield  {author} {\bibinfo {author} {\bibfnamefont {J.~R.~M.}\
  \bibnamefont {Howell}}\ and\ \bibinfo {author} {\bibfnamefont {M.~P.~S.}\
  \bibnamefont {Robert}},\ }\href@noop {} {\emph {\bibinfo {title} {Thermal
  radiation heat transfer}}},\ \bibinfo {edition} {sixth}\ ed.\ (\bibinfo
  {publisher} {CRC Press},\ \bibinfo {address} {Boca Raton},\ \bibinfo {year}
  {2016})\BibitemShut {NoStop}%
\bibitem [{\citenamefont {Wang}\ and\ \citenamefont
  {Jacques}(1992)}]{wang_monte_1992}%
  \BibitemOpen
  \bibfield  {author} {\bibinfo {author} {\bibfnamefont {L.}~\bibnamefont
  {Wang}}\ and\ \bibinfo {author} {\bibfnamefont {S.~L.}\ \bibnamefont
  {Jacques}},\ }\bibfield  {title} {\emph {\bibinfo {title} {Monte {Carlo}
  {Modeling} of {Light} {Transport} in {Multi}-layered {Tissues} in {Standard}
  {C}}},\ }\href {http://coilab.caltech.edu/mcr5/Mcman.pdf} {\bibfield
  {journal} {\bibinfo  {journal} {University of Texas}\ ,\ \bibinfo {pages}
  {1}} (\bibinfo {year} {1992})}\BibitemShut {NoStop}%
\bibitem [{\citenamefont {Redmond}\ \emph {et~al.}(2004)\citenamefont
  {Redmond}, \citenamefont {Rand}, \citenamefont {Ruan},\ and\ \citenamefont
  {Kaviany}}]{redmond_multiple_2004}%
  \BibitemOpen
  \bibfield  {author} {\bibinfo {author} {\bibfnamefont {S.}~\bibnamefont
  {Redmond}}, \bibinfo {author} {\bibfnamefont {S.~C.}\ \bibnamefont {Rand}},
  \bibinfo {author} {\bibfnamefont {X.~L.}\ \bibnamefont {Ruan}},\ and\
  \bibinfo {author} {\bibfnamefont {M.}~\bibnamefont {Kaviany}},\ }\bibfield
  {title} {\emph {\bibinfo {title} {Multiple scattering and nonlinear thermal
  emission of {Yb}$^{\textrm{3+}}$,
  {Er}$^{\textrm{3+}}$:{Y$_{\textrm{2}}$O$_{\textrm{3}}$} nanopowders}},\
  }\href {https://doi.org/10.1063/1.1667274} {\bibfield  {journal} {\bibinfo
  {journal} {Journal of Applied Physics}\ }\textbf {\bibinfo {volume} {95}},\
  \bibinfo {pages} {4069} (\bibinfo {year} {2004})}\BibitemShut {NoStop}%
\bibitem [{\citenamefont {Zhang}(2007)}]{zhang_nanomicroscale_2007}%
  \BibitemOpen
  \bibfield  {author} {\bibinfo {author} {\bibfnamefont {Z.~M.}\ \bibnamefont
  {Zhang}},\ }\href@noop {} {\emph {\bibinfo {title} {Nano/microscale heat
  transfer}}}\ (\bibinfo  {publisher} {McGraw-Hill},\ \bibinfo {address} {New
  York},\ \bibinfo {year} {2007})\BibitemShut {NoStop}%
\bibitem [{\citenamefont {Gajdo\v{s}}\ \emph {et~al.}(2006)\citenamefont
  {Gajdo\v{s}}, \citenamefont {Hummer}, \citenamefont {Kresse}, \citenamefont
  {Furthm\"{u}ller},\ and\ \citenamefont {Bechstedt}}]{gajdos_linear_2006}%
  \BibitemOpen
  \bibfield  {author} {\bibinfo {author} {\bibfnamefont {M.}~\bibnamefont
  {Gajdo\v{s}}}, \bibinfo {author} {\bibfnamefont {K.}~\bibnamefont {Hummer}},
  \bibinfo {author} {\bibfnamefont {G.}~\bibnamefont {Kresse}}, \bibinfo
  {author} {\bibfnamefont {J.}~\bibnamefont {Furthm\"{u}ller}},\ and\ \bibinfo
  {author} {\bibfnamefont {F.}~\bibnamefont {Bechstedt}},\ }\bibfield  {title}
  {\emph {\bibinfo {title} {Linear optical properties in the
  projector-augmented wave methodology}},\ }\href
  {https://doi.org/10.1103/PhysRevB.73.045112} {\bibfield  {journal} {\bibinfo
  {journal} {Physical Review B}\ }\textbf {\bibinfo {volume} {73}},\ \bibinfo
  {pages} {045112} (\bibinfo {year} {2006})}\BibitemShut {NoStop}%
\bibitem [{\citenamefont {Barker}(1964)}]{barker_transverse_1964}%
  \BibitemOpen
  \bibfield  {author} {\bibinfo {author} {\bibfnamefont {A.~S.}\ \bibnamefont
  {Barker}},\ }\bibfield  {title} {\emph {\bibinfo {title} {Transverse and
  {Longitudinal} {Optic} {Mode} {Study} in {Mg}{F}$_{\textrm{2}}$ and
  {Zn}{F}$_{\textrm{2}}$}},\ }\href {https://doi.org/10.1103/PhysRev.136.A1290}
  {\bibfield  {journal} {\bibinfo  {journal} {Physical Review}\ }\textbf
  {\bibinfo {volume} {136}},\ \bibinfo {pages} {A1290} (\bibinfo {year}
  {1964})}\BibitemShut {NoStop}%
\bibitem [{\citenamefont
  {Schubert}(2016)}]{schubert_coordinate-invariant_2016}%
  \BibitemOpen
  \bibfield  {author} {\bibinfo {author} {\bibfnamefont {M.}~\bibnamefont
  {Schubert}},\ }\bibfield  {title} {\emph {\bibinfo {title}
  {Coordinate-{Invariant} {Lyddane}-{Sachs}-{Teller} {Relationship} for {Polar}
  {Vibrations} in {Materials} with {Monoclinic} and {Triclinic} {Crystal}
  {Systems}}},\ }\href
  {https://link.aps.org/doi/10.1103/PhysRevLett.117.215502} {\bibfield
  {journal} {\bibinfo  {journal} {Physical Review Letters}\ }\textbf {\bibinfo
  {volume} {117}} (\bibinfo {year} {2016})}\BibitemShut {NoStop}%
\bibitem [{\citenamefont {Kresse}\ and\ \citenamefont
  {Joubert}(1999)}]{kresse_ultrasoft_1999}%
  \BibitemOpen
  \bibfield  {author} {\bibinfo {author} {\bibfnamefont {G.}~\bibnamefont
  {Kresse}}\ and\ \bibinfo {author} {\bibfnamefont {D.}~\bibnamefont
  {Joubert}},\ }\bibfield  {title} {\emph {\bibinfo {title} {From ultrasoft
  pseudopotentials to the projector augmented-wave method}},\ }\href
  {https://doi.org/10.1103/PhysRevB.59.1758} {\bibfield  {journal} {\bibinfo
  {journal} {Physical Review B}\ }\textbf {\bibinfo {volume} {59}},\ \bibinfo
  {pages} {1758} (\bibinfo {year} {1999})}\BibitemShut {NoStop}%
\bibitem [{\citenamefont {Kresse}\ and\ \citenamefont
  {Hafner}(1993)}]{kresse_ab_1993}%
  \BibitemOpen
  \bibfield  {author} {\bibinfo {author} {\bibfnamefont {G.}~\bibnamefont
  {Kresse}}\ and\ \bibinfo {author} {\bibfnamefont {J.}~\bibnamefont
  {Hafner}},\ }\bibfield  {title} {\emph {\bibinfo {title} {\textit{{Ab}
  initio} molecular dynamics for liquid metals}},\ }\href
  {https://doi.org/10.1103/PhysRevB.47.558} {\bibfield  {journal} {\bibinfo
  {journal} {Physical Review B}\ }\textbf {\bibinfo {volume} {47}},\ \bibinfo
  {pages} {558} (\bibinfo {year} {1993})}\BibitemShut {NoStop}%
\bibitem [{\citenamefont {Strauch}\ and\ \citenamefont
  {Dorner}(1993)}]{strauch_lattice_1993}%
  \BibitemOpen
  \bibfield  {author} {\bibinfo {author} {\bibfnamefont {D.}~\bibnamefont
  {Strauch}}\ and\ \bibinfo {author} {\bibfnamefont {B.}~\bibnamefont
  {Dorner}},\ }\bibfield  {title} {\emph {\bibinfo {title} {Lattice dynamics of
  alpha -quartz. {I}. {Experiment}}},\ }\href
  {https://doi.org/10.1088/0953-8984/5/34/003} {\bibfield  {journal} {\bibinfo
  {journal} {Journal of Physics: Condensed Matter}\ }\textbf {\bibinfo {volume}
  {5}},\ \bibinfo {pages} {6149} (\bibinfo {year} {1993})}\BibitemShut
  {NoStop}%
\bibitem [{\citenamefont {Perdew}\ \emph {et~al.}(1996)\citenamefont {Perdew},
  \citenamefont {Burke},\ and\ \citenamefont
  {Ernzerhof}}]{perdew_generalized_1996}%
  \BibitemOpen
  \bibfield  {author} {\bibinfo {author} {\bibfnamefont {J.~P.}\ \bibnamefont
  {Perdew}}, \bibinfo {author} {\bibfnamefont {K.}~\bibnamefont {Burke}},\ and\
  \bibinfo {author} {\bibfnamefont {M.}~\bibnamefont {Ernzerhof}},\ }\bibfield
  {title} {\emph {\bibinfo {title} {Generalized {Gradient} {Approximation}
  {Made} {Simple}}},\ }\href {https://doi.org/10.1103/PhysRevLett.77.3865}
  {\bibfield  {journal} {\bibinfo  {journal} {Physical Review Letters}\
  }\textbf {\bibinfo {volume} {77}},\ \bibinfo {pages} {3865} (\bibinfo {year}
  {1996})}\BibitemShut {NoStop}%
\bibitem [{\citenamefont {Alkauskas}\ \emph {et~al.}(2008)\citenamefont
  {Alkauskas}, \citenamefont {Broqvist},\ and\ \citenamefont
  {Pasquarello}}]{alkauskas_defect_2008}%
  \BibitemOpen
  \bibfield  {author} {\bibinfo {author} {\bibfnamefont {A.}~\bibnamefont
  {Alkauskas}}, \bibinfo {author} {\bibfnamefont {P.}~\bibnamefont
  {Broqvist}},\ and\ \bibinfo {author} {\bibfnamefont {A.}~\bibnamefont
  {Pasquarello}},\ }\bibfield  {title} {\emph {\bibinfo {title} {Defect
  {Energy} {Levels} in {Density} {Functional} {Calculations}: {Alignment} and
  {Band} {Gap} {Problem}}},\ }\href
  {https://doi.org/10.1103/PhysRevLett.101.046405} {\bibfield  {journal}
  {\bibinfo  {journal} {Physical Review Letters}\ }\textbf {\bibinfo {volume}
  {101}},\ \bibinfo {pages} {046405} (\bibinfo {year} {2008})}\BibitemShut
  {NoStop}%
\bibitem [{\citenamefont {Korabel'nikov}\ and\ \citenamefont
  {Zhuravlev}(2018)}]{korabelnikov_structural_2018}%
  \BibitemOpen
  \bibfield  {author} {\bibinfo {author} {\bibfnamefont {D.}~\bibnamefont
  {Korabel'nikov}}\ and\ \bibinfo {author} {\bibfnamefont {Y.}~\bibnamefont
  {Zhuravlev}},\ }\bibfield  {title} {\emph {\bibinfo {title} {Structural,
  elastic, electronic and vibrational properties of a series of sulfates from
  first principles calculations}},\ }\href
  {https://doi.org/10.1016/j.jpcs.2018.03.037} {\bibfield  {journal} {\bibinfo
  {journal} {Journal of Physics Chemistry of Solids}\ }\textbf {\bibinfo
  {volume} {119}},\ \bibinfo {pages} {114} (\bibinfo {year}
  {2018})}\BibitemShut {NoStop}%
\bibitem [{\citenamefont {Amiryan}\ \emph {et~al.}(1977)\citenamefont
  {Amiryan}, \citenamefont {Gurvieh}, \citenamefont {Katomina}, \citenamefont
  {Petrova}, \citenamefont {Soshchin},\ and\ \citenamefont
  {Tombak}}]{amiryan_recombination_1977}%
  \BibitemOpen
  \bibfield  {author} {\bibinfo {author} {\bibfnamefont {A.~M.}\ \bibnamefont
  {Amiryan}}, \bibinfo {author} {\bibfnamefont {A.~M.}\ \bibnamefont
  {Gurvieh}}, \bibinfo {author} {\bibfnamefont {R.~V.}\ \bibnamefont
  {Katomina}}, \bibinfo {author} {\bibfnamefont {I.~Y.}\ \bibnamefont
  {Petrova}}, \bibinfo {author} {\bibfnamefont {N.~P.}\ \bibnamefont
  {Soshchin}},\ and\ \bibinfo {author} {\bibfnamefont {M.~I.}\ \bibnamefont
  {Tombak}},\ }\bibfield  {title} {\emph {\bibinfo {title} {Recombination
  processes and emission spectrum of terbium in oxysulfides}},\ }\href
  {https://doi.org/10.1007/BF00625902} {\bibfield  {journal} {\bibinfo
  {journal} {Journal of Applied Spectroscopy}\ }\textbf {\bibinfo {volume}
  {27}},\ \bibinfo {pages} {1159} (\bibinfo {year} {1977})}\BibitemShut
  {NoStop}%
\bibitem [{\citenamefont {Kroumova}\ \emph {et~al.}(2003)\citenamefont
  {Kroumova}, \citenamefont {Aroyo}, \citenamefont {Perez-Mato}, \citenamefont
  {Kirov}, \citenamefont {Capillas}, \citenamefont {Ivantchev},\ and\
  \citenamefont {Wondratschek}}]{kroumova_bilbao_2003}%
  \BibitemOpen
  \bibfield  {author} {\bibinfo {author} {\bibfnamefont {E.}~\bibnamefont
  {Kroumova}}, \bibinfo {author} {\bibfnamefont {M.}~\bibnamefont {Aroyo}},
  \bibinfo {author} {\bibfnamefont {J.}~\bibnamefont {Perez-Mato}}, \bibinfo
  {author} {\bibfnamefont {A.}~\bibnamefont {Kirov}}, \bibinfo {author}
  {\bibfnamefont {C.}~\bibnamefont {Capillas}}, \bibinfo {author}
  {\bibfnamefont {S.}~\bibnamefont {Ivantchev}},\ and\ \bibinfo {author}
  {\bibfnamefont {H.}~\bibnamefont {Wondratschek}},\ }\bibfield  {title} {\emph
  {\bibinfo {title} {Bilbao {Crystallographic} {Server} : {Useful} {Databases}
  and {Tools} for {Phase}-{Transition} {Studies}}},\ }\href
  {https://doi.org/10.1080/0141159031000076110} {\bibfield  {journal} {\bibinfo
   {journal} {Phase Transitions}\ }\textbf {\bibinfo {volume} {76}},\ \bibinfo
  {pages} {155} (\bibinfo {year} {2003})}\BibitemShut {NoStop}%
\bibitem [{\citenamefont {Glazer}(2009)}]{glazer_vibrate_2009}%
  \BibitemOpen
  \bibfield  {author} {\bibinfo {author} {\bibfnamefont {A.~M.}\ \bibnamefont
  {Glazer}},\ }\bibfield  {title} {\emph {\bibinfo {title} {\textit{{VIBRATE}!}
  {A} program to compute irreducible representations for atomic vibrations in
  crystals}},\ }\href {https://doi.org/10.1107/S0021889809040424} {\bibfield
  {journal} {\bibinfo  {journal} {Journal of Applied Crystallography}\ }\textbf
  {\bibinfo {volume} {42}},\ \bibinfo {pages} {1194} (\bibinfo {year}
  {2009})}\BibitemShut {NoStop}%
\bibitem [{\citenamefont {Philipp}(1966)}]{philipp_optical_1966}%
  \BibitemOpen
  \bibfield  {author} {\bibinfo {author} {\bibfnamefont {H.~R.}\ \bibnamefont
  {Philipp}},\ }\bibfield  {title} {\emph {\bibinfo {title} {Optical
  transitions in crystalline and fused quartz}},\ }\href
  {https://doi.org/10.1016/0038-1098(66)90109-8} {\bibfield  {journal}
  {\bibinfo  {journal} {Solid State Communications}\ }\textbf {\bibinfo
  {volume} {4}},\ \bibinfo {pages} {73} (\bibinfo {year} {1966})}\BibitemShut
  {NoStop}%
\bibitem [{\citenamefont {Gervais}\ and\ \citenamefont
  {Piriou}(1975)}]{gervais_temperature_1975}%
  \BibitemOpen
  \bibfield  {author} {\bibinfo {author} {\bibfnamefont {F.}~\bibnamefont
  {Gervais}}\ and\ \bibinfo {author} {\bibfnamefont {B.}~\bibnamefont
  {Piriou}},\ }\bibfield  {title} {\emph {\bibinfo {title} {Temperature
  dependence of transverse and longitudinal optic modes in the $\alpha$ and
  $\beta$ phases of quartz}},\ }\href
  {https://doi.org/10.1103/PhysRevB.11.3944} {\bibfield  {journal} {\bibinfo
  {journal} {Physical Review B}\ }\textbf {\bibinfo {volume} {11}},\ \bibinfo
  {pages} {3944} (\bibinfo {year} {1975})}\BibitemShut {NoStop}%
\bibitem [{\citenamefont {Khashan}\ and\ \citenamefont
  {Nassif}(2001)}]{khashan_dispersion_2001}%
  \BibitemOpen
  \bibfield  {author} {\bibinfo {author} {\bibfnamefont {M.~A.}\ \bibnamefont
  {Khashan}}\ and\ \bibinfo {author} {\bibfnamefont {A.~Y.}\ \bibnamefont
  {Nassif}},\ }\bibfield  {title} {\emph {\bibinfo {title} {Dispersion of the
  optical constants of quartz and polymethyl methacrylate glasses in a wide
  spectral range: 0.2–3 μm}},\ }\href
  {https://doi.org/10.1016/S0030-4018(00)01152-4} {\bibfield  {journal}
  {\bibinfo  {journal} {Optics Communications}\ }\textbf {\bibinfo {volume}
  {188}},\ \bibinfo {pages} {129} (\bibinfo {year} {2001})}\BibitemShut
  {NoStop}%
\bibitem [{\citenamefont {Ravindra}\ \emph {et~al.}(2007)\citenamefont
  {Ravindra}, \citenamefont {Ganapathy},\ and\ \citenamefont
  {Choi}}]{ravindra_energy_2007}%
  \BibitemOpen
  \bibfield  {author} {\bibinfo {author} {\bibfnamefont {N.}~\bibnamefont
  {Ravindra}}, \bibinfo {author} {\bibfnamefont {P.}~\bibnamefont
  {Ganapathy}},\ and\ \bibinfo {author} {\bibfnamefont {J.}~\bibnamefont
  {Choi}},\ }\bibfield  {title} {\emph {\bibinfo {title} {Energy
  gap–refractive index relations in {semiconductors} - {An} overview}},\
  }\href {https://doi.org/10.1016/j.infrared.2006.04.001} {\bibfield  {journal}
  {\bibinfo  {journal} {Infrared Physics \& Technology}\ }\textbf {\bibinfo
  {volume} {50}},\ \bibinfo {pages} {21} (\bibinfo {year} {2007})}\BibitemShut
  {NoStop}%
\bibitem [{\citenamefont {Adachi}(1999)}]{adachi_optical_1999}%
  \BibitemOpen
  \bibfield  {author} {\bibinfo {author} {\bibfnamefont {S.}~\bibnamefont
  {Adachi}},\ }\href@noop {} {\emph {\bibinfo {title} {Optical {Properties} of
  {Crystalline} and {Amorphous} {Semiconductors}}}}\ (\bibinfo  {publisher}
  {Springer US},\ \bibinfo {address} {Boston},\ \bibinfo {year}
  {1999})\BibitemShut {NoStop}%
\bibitem [{\citenamefont {{ASTM G159-98}}(1998)}]{astm_g159-98_standard_1998}%
  \BibitemOpen
  \bibfield  {author} {\bibinfo {author} {\bibnamefont {{ASTM G159-98}}},\
  }\href {http://www.astm.org/cgi-bin/resolver.cgi?G159-98} {\emph {\bibinfo
  {title} {Standard {Tables} for {Solar} {Spectral}}}},\ \bibinfo {type} {Tech.
  Rep.}\ (\bibinfo  {institution} {ASTM International},\ \bibinfo {year}
  {1998})\BibitemShut {NoStop}%
\bibitem [{\citenamefont {Lord}(1992)}]{lord_new_1992}%
  \BibitemOpen
  \bibfield  {author} {\bibinfo {author} {\bibfnamefont {S.~D.}\ \bibnamefont
  {Lord}},\ }\href {https://ntrs.nasa.gov/search.jsp?R=19930010877} {\emph
  {\bibinfo {title} {A software tool for computing transmission}}},\ \bibinfo
  {type} {Tech. Rep.}\ (\bibinfo {year} {1992})\BibitemShut {NoStop}%
\bibitem [{\citenamefont {Bohren}\ and\ \citenamefont
  {Huffman}(1983)}]{bohren_absorption_1983}%
  \BibitemOpen
  \bibfield  {author} {\bibinfo {author} {\bibfnamefont {C.~F.}\ \bibnamefont
  {Bohren}}\ and\ \bibinfo {author} {\bibfnamefont {D.~R.}\ \bibnamefont
  {Huffman}},\ }\href@noop {} {\emph {\bibinfo {title} {Absorption and
  scattering of light by small particles}}}\ (\bibinfo  {publisher} {Wiley},\
  \bibinfo {address} {New York},\ \bibinfo {year} {1983})\BibitemShut {NoStop}%
\end{thebibliography}%

\end{document}